\pdfoutput=1
\documentclass[preprint2]{aastex}

\slugcomment{ANL-HEP-PR-12-60; FERMILAB-PUB-12-393-AE-CD-PPD}

\shorttitle{PreCam}
\shortauthors{Kuehn, K. et al.}

\begin{document}

\title{PreCam, a Precursor Observational Campaign\\
for Calibration of the Dark Energy Survey}

\author{K. Kuehn\altaffilmark{1,2}, 
          S. Kuhlmann\altaffilmark{1,2},
          S. Allam\altaffilmark{3},
          J. T. Annis\altaffilmark{3},
          T. Bailey\altaffilmark{1,4},
          E. Balbinot\altaffilmark{5,6},
          J. P. Bernstein\altaffilmark{1},
          T. Biesiadzinski\altaffilmark{7},
          D. L. Burke\altaffilmark{8},
          M. Butner\altaffilmark{9},
          J. I. B. Camargo\altaffilmark{6,10}, 
          L. A. N. da Costa\altaffilmark{6,10}, 
          D. DePoy\altaffilmark{11},
          H. T. Diehl\altaffilmark{3}, 
          J. P. Dietrich\altaffilmark{7}, J. Estrada\altaffilmark{3}, A. Fausti\altaffilmark{6},
          B. Gerke\altaffilmark{8,12}, V. Guarino\altaffilmark{1},
          H. H. Head\altaffilmark{9},
          R. Kessler\altaffilmark{13},
          H. Lin\altaffilmark{3},
          W. Lorenzon\altaffilmark{7},
          M. A. G. Maia\altaffilmark{6,10},
          L. Maki\altaffilmark{7,14},
          J. Marshall\altaffilmark{10},
          B. Nord\altaffilmark{7},
          E. Neilsen\altaffilmark{3},
          R. L. C. Ogando\altaffilmark{6,10},
          D. Park\altaffilmark{3,15},
          J. Peoples\altaffilmark{3},
          D. Rastawicki\altaffilmark{16},
          J.-P. Rheault\altaffilmark{10}, B. Santiago\altaffilmark{5,6},
          M. Schubnell\altaffilmark{7}, P. Seitzer\altaffilmark{17},
          J. A. Smith\altaffilmark{9}, H. Spinka\altaffilmark{1},
          A. Sypniewski\altaffilmark{7},
          G. Tarle\altaffilmark{7},
          D. L. Tucker\altaffilmark{3,2},
          A. Walker\altaffilmark{18},
          W. Wester\altaffilmark{3}\\
(the Dark Energy Survey Collaboration)
\\(ANL-HEP-PR-12-60; FERMILAB-PUB-12-393-AE-CD-PPD)
}

\altaffiltext{1}{High Energy Physics Division, Argonne National Laboratory, Lemont, IL 60439}
\altaffiltext{2}{Corresponding Authors; email: kkuehn@anl.gov (KK), kuhlmann@anl.gov (SK), dtucker@fnal.gov (DLT)}
\altaffiltext{3}{Fermi National Accelerator Laboratory, Batavia, IL 60510}
\altaffiltext{4}{Princeton University, Princeton, NJ 08544}
\altaffiltext{5}{Instituto de F\'\i sica, UFRGS, Porto Alegre, RS - 91501-970, Brazil}
\altaffiltext{6}{Laborat\'orio Interinstitucional de e-Astronomia -- LIneA, Rio de Janeiro, RJ - 20921-400, Brazil}
\altaffiltext{7}{Department of Physics, University of Michigan, Ann Arbor, MI 48109}
\altaffiltext{8}{SLAC National Accelerator Laboratory, Menlo Park, CA 94025}
\altaffiltext{9}{Department of Physics and Astronomy, Austin Peay State University, Clarksville, TN 37044}
\altaffiltext{10}{Observat\'orio Nacional, Rio de Janeiro, RJ - 20921-400, Brazil}
\altaffiltext{11}{Department of Physics \& Astronomy, Texas A\&M University, College Station, TX 77843}
\altaffiltext{12}{Lawrence Berkeley National Laboratory, Berkeley, CA 94720}
\altaffiltext{13}{Department of Astronomy and Astrophysics, University of Chicago, Chicago, IL 60637}
\altaffiltext{14}{Department of Physics and Astronomy, Wayne State University, Detroit, MI 48202}
\altaffiltext{15}{Illinois Math and Science Academy, Aurora, IL 60506}
\altaffiltext{16}{Department of Physics, Stanford University, Stanford, CA 94305}
\altaffiltext{17}{Department of Astronomy, University of Michigan, Ann Arbor, MI 48109}
\altaffiltext{18}{Cerro Tololo Interamerican Observatory, La Serena, Chile}

\begin{abstract}
PreCam, a precursor observational campaign supporting the Dark Energy Survey (DES), is designed to produce a photometric and astrometric catalog of nearly a hundred thousand standard stars within the DES footprint, while the PreCam instrument also serves as a prototype testbed for the Dark Energy Camera (DECam)'s hardware and software.  This catalog represents a potential 100-fold increase in Southern Hemisphere photometric standard stars, and therefore will be an important component in the calibration of the Dark Energy Survey.  We provide details on the PreCam instrument's design, construction and testing, as well as results from a subset of the 51 nights of PreCam survey observations on the University of Michigan Department of Astronomy's Curtis-Schmidt telescope at Cerro Tololo Inter-American Observatory.  We briefly describe the preliminary data processing pipeline that has been developed for PreCam data and the preliminary results of the instrument performance, as well as astrometry and photometry of a sample of stars previously included in other southern sky surveys.
\end{abstract}


\section{The Dark Energy Survey in the Context of Current Cosmology}

The Dark Energy Survey, or DES, will map 5000 deg$^2$ of the southern galactic cap to observe more than 10$^8$ faint galaxies, more than 3000 Type Ia supernovae, and myriad other objects, to determine the nature and temporal evolution (if any) of dark energy \citep{Annis, Joe}.  The DES utilizes the Dark Energy Camera, or DECam \citep{Brenna}, installed on the Blanco telescope at Cerro Tololo Interamerican Observatory, to observe each pointing numerous times in each in five different passbands: g, r, i, z, and Y (the first four of which are similar, but not identical, to the Sloan Digital Sky Survey, or SDSS \citep{Sloan} filters).  The DECam includes 62 2k x 4k pixel science CCDs (along with additional guide and focus/alignment CCDs) totaling $\sim$570 megapixels, creating a field of view of $\sim$3 deg$^2$.   A distinguishing characteristic of DECam is the choice of 200-micron thick, fully-depleted, n-type, red-sensitive CCDs for the imager array. These CCDs have a much better red-sensitivity, with an efficiency of greater than 50\% per photon at wavelength 1000 nm, nearly an order of magnitude better than standard thin CCDs \citep{MegaCam} at that wavelength.
  
%% Figure~\ref{f1} shows the baseline footprint of the Dark Energy Survey and the dust associated with the galactic plane \citep{Schlegel}, which we avoid since we aim mainly at extragalactic sources.  Figure~\ref{f2} shows the arrangement of DES CCDs with the image of the full moon to scale on the focal plane (though of course we will not observe the moon directly with DECam).

The goal of the DES is a factor of 3-5 improvement in the measured Dark Energy Task Force Figure of Merit of the dark energy parameters $w_0$ and $w_a$ over a Stage II experiment \citep{DETF}.  To achieve this goal, the DES has an all-sky photometric calibration requirement of 2\% (and a goal of 1\%).  Because the science requirements for the DES are quite stringent, the characteristics of the DECam must likewise be precisely controlled.  One important method of calibrating the performance of the CCDs (and the entire optical system) during the Survey is to compare DECam measurements to previous measurements of known standard stars.  Several catalogs of equatorial and Southern Hemisphere standard stars exist \citep{Sloan2, South, South2}, but the extremely sparse nature of the observations render them inadequate for the needs of the DES.  Furthermore, there are almost no standard stars that have been observed in the Y band.  Thus, the efforts to provide calibration standards for the DECam are vital to the success of the overall Survey.  The PreCam instrument contains two CCDs taken from among the spare devices made for the DECam itself, making the focal plane similar to a 1/31st-scale version of the DECam focal plane.  This allows the PreCam survey observations to provide many of the necessary calibration standard stars for the DES, especially in the early years of the Survey.  This in turn potentially allows up to 10\% more of the DES time to be dedicated to science observations rather than calibration measurements, while still maintaining the design requirement of 2\% photometric accuracy throughout the survey area.  Much of the PreCam instrument's hardware and software (such as the readout electronics) is likewise similar to that of the DECam, and has also been used for testing the DECam CCD electronics during the reassembly of the DECam system at CTIO after shipment.

\begin{figure*}
\epsscale{1.75}
%% \plotone{f1.eps}
%% \plotone{footgrid.eps}
\plotone{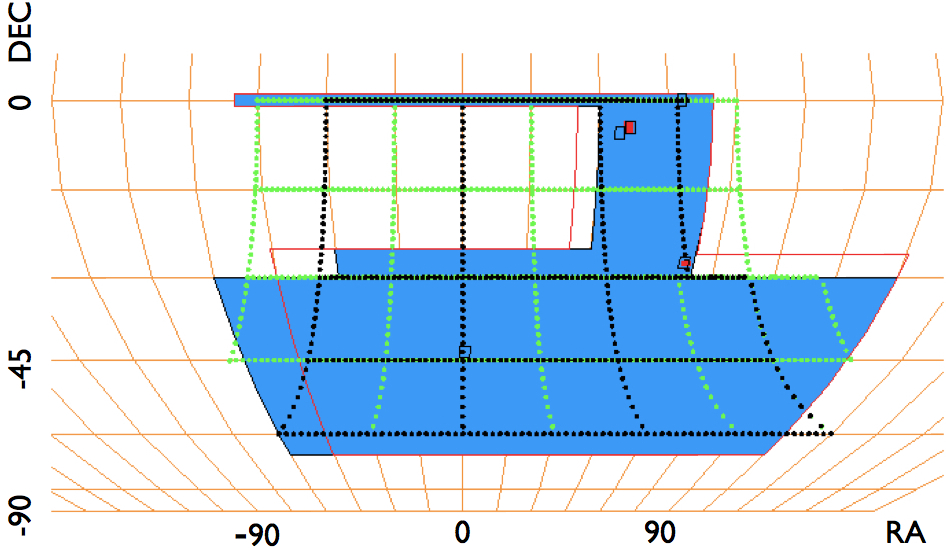}
\caption{The PreCam survey grid (black and green points), overlaid on an earlier proposed version of the DES footprint (blue shaded region with SN fields as small boxes; the most recent footprint has since undergone minor revisions).  The grid facilitates the connection of the region of overlap with the SDSS data to the region of overlap with the VHS and SPT data.  During standard DES operations, the camera will intersect one of these grid points approximately every 20 minutes throughout the night.}
\label{f3}
\end{figure*}

To accomplish its calibration goals, the PreCam survey observes relatively bright (magnitude 14-18) stars predominantly within the DES footprint using filters very similar to the DES grizY filter system \citep{Filter}.  For this survey, the PreCam instrument is mounted on the University of Michigan Department of Astronomy's Curtis-Schmidt (C-S) telescope at CTIO. The first major step of the PreCam survey is to observe thousands of existing standard stars that have been catalogued as part of the SDSS or u$^{\prime}$g$^{\prime}$r$^{\prime}$i$^{\prime}$z$^{\prime}$ standard star surveys \citep{Sloan2, South, South2}. We then determine the survey's photometric accuracy by comparing our observations to these other catalogs.  We then use further PreCam observations of target candidate standard stars throughout the observed survey grid ($\sim$10$^{5}$ stars overall) to provide a significantly expanded catalog that will be used to improve the overall photometric accuracy of the DES. On the C-S, the PreCam instrument's field-of-view is 1.6$^{\circ}$ $\times$ 1.6$^{\circ}$ (or 2.56 deg$^2$).  The PreCam survey was designed to cover 10\% of the DES footprint in a sparse grid in RA and DEC, and each field in this grid was planned to be observed several times in each of the five filters.  This grid pattern ensures that the DECam will regularly observe these calibrated standards during routine survey operations, thus obviating much of the need for dedicated standard star observations with DECam itself and thereby saving valuable observing time that can be devoted to further DES science goals. Figure~\ref{f3} shows the planned PreCam survey grid overlaid on the DES footprint.  In Section 2 we describe the design and construction of the PreCam instrument, as well as preliminary tests of its performance.  In Section 3, we describe PreCam survey operations, while Section 4 describes the preliminary post-survey processing of the data.  Sections 5 covers the results of this analysis, and in Section 6 we discuss the impact of these results in the context of DES calibrations.  In Section 7, we conclude with a look toward the future prospects for PreCam.

\section{Instrumentation and Bench Tests}

\begin{figure*}
\epsscale{2.5}
\plottwo{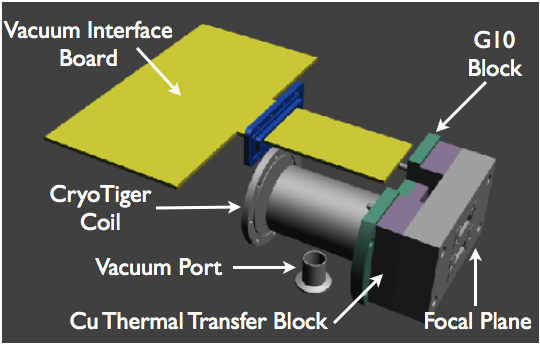}{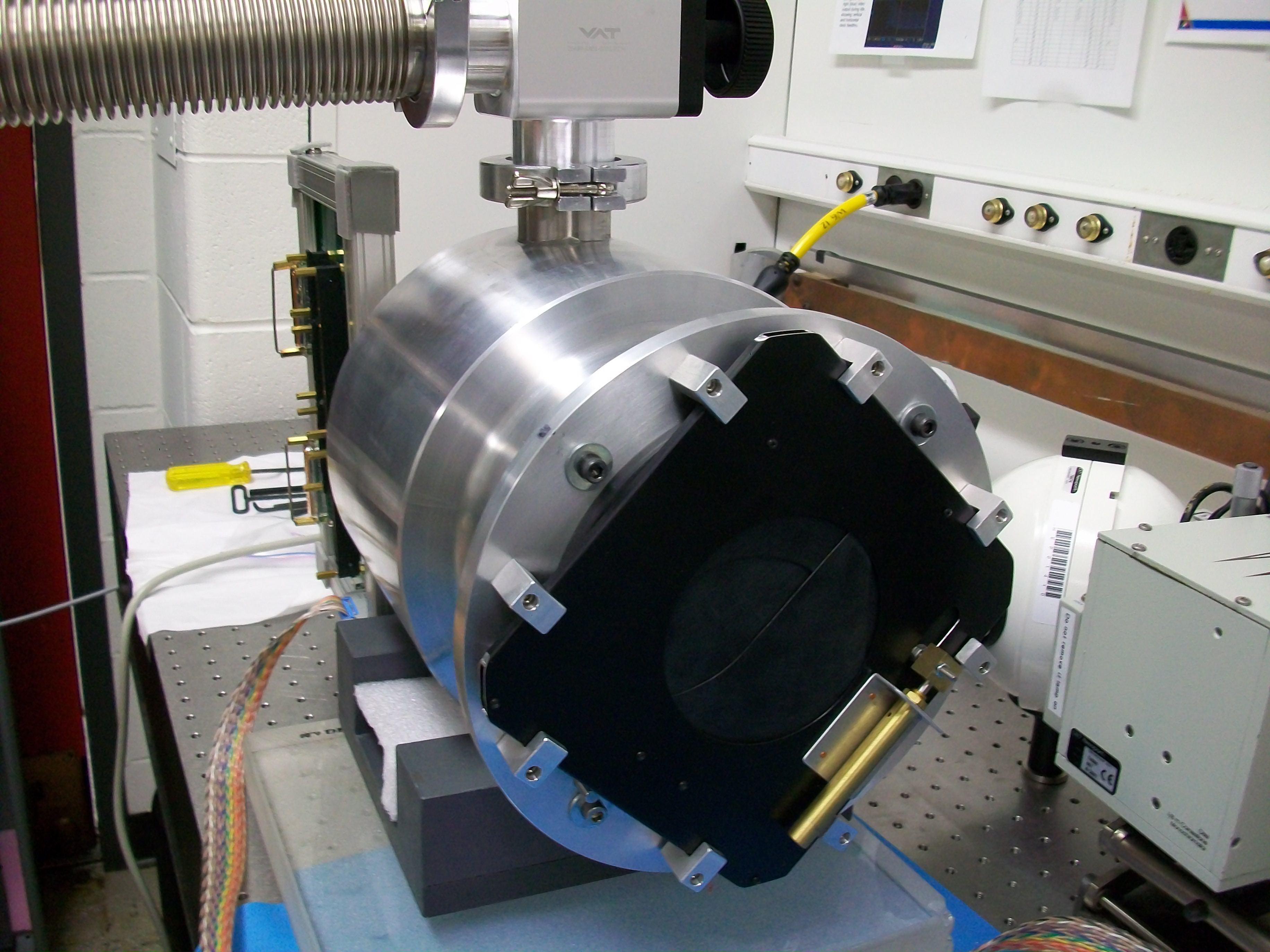}
%% \plottwo{ccd_qe.eps}{fov_moon.eps}
\caption{Left: A CAD rendering of the PreCam instrument showing the various components (without the encasing vacuum vessel).  Right: Photograph of the PreCam instrument and the bench test setup at Argonne National Laboratory, with the shutter installed in front of the dewar window.}
\label{f5}
\end{figure*}

The PreCam instrument was primarily designed and built at Argonne National Laboratory, with the data acquisition (DAQ) electronics components contributed by Fermi National Accelerator Laboratory.  The CCD detectors reside in an aluminum pressure vessel designed to operate at $\sim$10$^{-6}$ mbar and -100$^{\circ}$ C, as required to reduce thermal noise to acceptable levels within the CCDs.  In addition to the throughputs for cryogenic cooling and vacuum, the vessel has a 4 inch diameter sapphire window with an anti-reflective coating which allowed $\sim$ 99$\%$ transmission of light from the telescope to the CCDs\footnote{Al$_2$O$_3$ dewar window with A/R coating manufactured by JML Optical Industries, LLC, Rochester, NY 14625 (http://jmloptical.com)}.  A custom-designed shutter\footnote{Compressed-gas actuated shutter manufactured by Packard Shutter Co., Fiddletown, CA, 95629 (http://www.packardshutter.com)} is mounted in front of the vessel window and is actuated by compressed gas with an opening/closing stroke lasting less than 0.25 seconds (with the precise time depending somewhat on the applied gas pressure).  Dry nitrogen is fed to the shutter and regulated by an electronically-controlled valve that receives the open signal in coincidence with the recording of the observed photoelectrons by the CCD.  The vessel also has a Vacuum Interface Board (VIB) for signal readout from the CCDs \citep{VIB}.  The CCDs themselves are mounted on an aluminum focal plane immediately behind the vessel window and are positioned with an accuracy of better than 1mm, with a spacing between the two CCDs of $\sim$0.5 mm.  Kapton cables 30 cm in length separately connect each CCD to the VIB.  The focal plane is secured to a Cu thermal transfer block, which provides thermal contact between the CCDs and the CryoTiger cold probe\footnote{PolyCold CryoTiger system manufactured by Brooks Automation, Inc., Chelmsford, MA 01824}.  A custom-designed block of G10 plastic secures the thermal transfer block to the outer aluminum vessel while minimizing thermal transfer to the vessel walls.  Two Pt thermocouples attached to the thermal transfer block are used to measure the temperature near the CCDs, and two 25 W heaters connected to the thermal transfer block complete the feedback loop for temperature control.  While the CryoTiger is designed to operate at full power continuously, a Lakeshore 332 (LS332) device monitors the thermocouples and provides power to the heaters to maintain temperature stability\footnote{LS332 manufactured by LakeShore Electronics, Westerville, OH 43226 (http://lakeshore.com)}.  To provide vacuum, a turbopump is attached to the vessel during testing and maintenance, but not operation\footnote{HiCUBE Turbomolecular pump manufactured by Pfeiffer Vacuum, GmbH, 35614 Asslar, Germany (http://www.pfeiffer-vacuum.com)}.  Additionally, a small quantity ($\sim$ 10 g) of charcoal getter is secured to the thermal transfer block to adsorb any gases not removed by the vacuum pump\footnote{Activated charcoal getter manufactured by Adsorbents \& Dessicants Corporation of America (ADCOA), Gardena, CA 90247 (http://www.adcoa.net)}.  After vacuum was established during bench tests, outgassing was determined to be minimal, and the presence of the getter allowed the operating pressure (10$^{-6}$ to 10$^{-5}$ mbar) to be maintained for more than a month without further pumping.  Outside the vessel, the VIB is connected to a Monsoon electronics crate containing pre-production DES clock and bias and data acquisition boards; the Monsoon crate is then connected to the control PC with a fiber optic S-Link interface \citep{FEE}.  The entire data DAQ system uses PANVIEW (a customized ``Pixel Acquisition Node'' LabVIEW interface), which is in turn controlled by the observer through a prototype of the Survey Image System Process Integration, or SISPI \citep{SISPI}, the software manager and user interface for the DECam.  See Figure~\ref{f5} for a CAD model of the PreCam instrument and for a photograph of the complete PreCam system, as well as Table~\ref{tab1} for the characteristics of the instrument.

\begin{table*}
%%\tablewidth{20pt}
\begin{center}
%%\centering
\begin{tabular}{lr}
\hline
\hline
CCD Type & 2 x DECam 2048 x 4086 pixels\\
CCD Pixel Scale & 1.45 arcsec/pixel\\
Digitization Precision & 16 bit/pixel \\
Number of Readout Ports & 4 (2 per CCD) \\
Window Type & 4 inch diameter Sapphire, AR Coated \\
Cooler Type and Setpoint & CryoTiger, -100$^{\circ}$ C \\
Readout Speed & $\sim$20 s \\
Gain & 4 ADU/photoelectron \\
Dark Noise & $\sim$1 ADU/min \\
Readout Noise & $\sim$1 ADU/pixel \\
Fullwell & 141K,113K,134K,144K e$^{-}$/pixel \\
Pixel Size & 15 $\mu$m x 15 $\mu$m \\
Field of View & 1.6$^{\circ}$ x 1.6$^{\circ}$ ($<$10\% vignetting in central 1.0$^{\circ}$ diameter area) \\
\hline
\end{tabular}
\end{center}
\caption{PreCam Instrument Characteristics}
\label{tab1}
\end{table*}

\begin{figure*}
\epsscale{2.5}
\plottwo{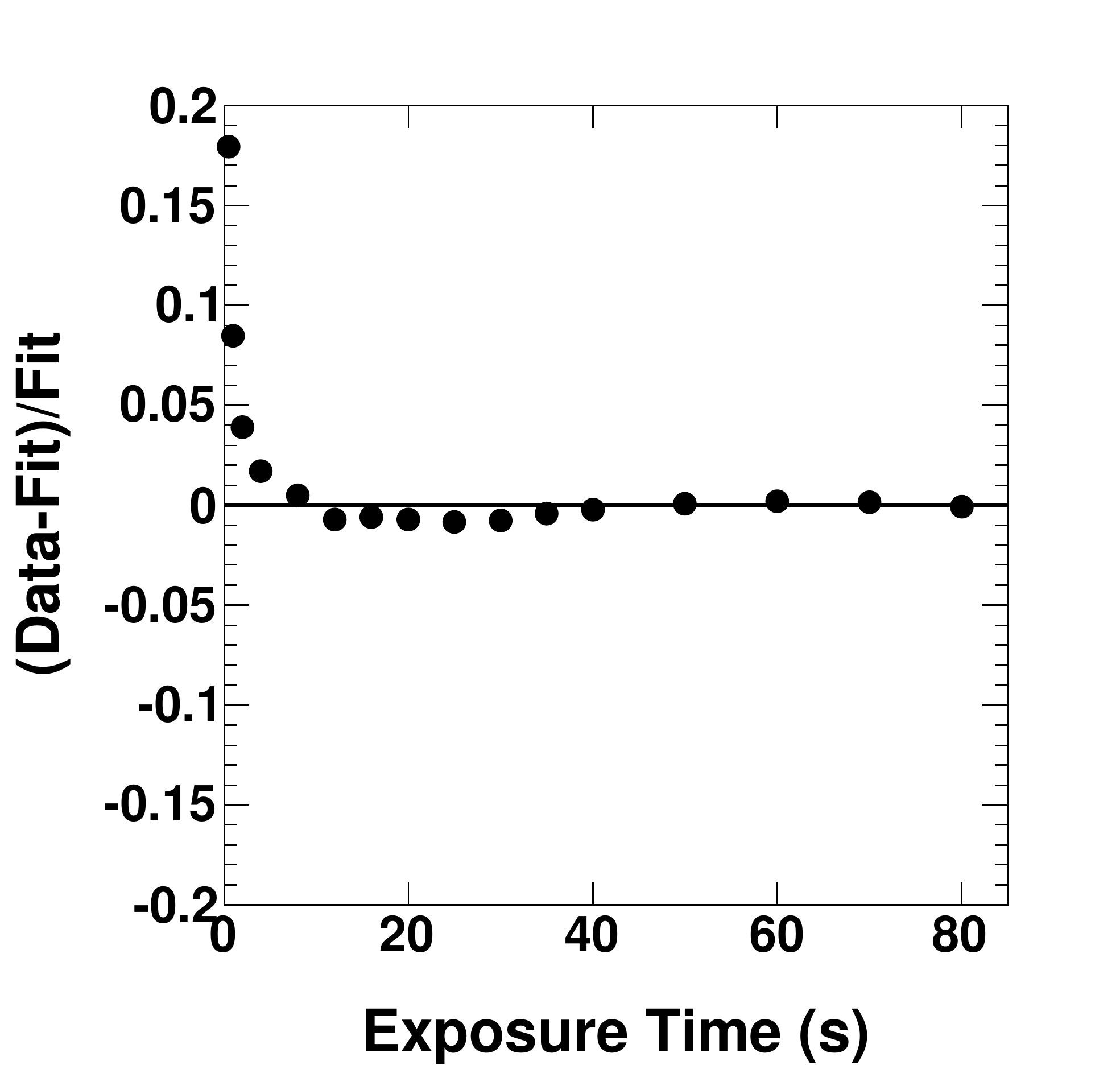}{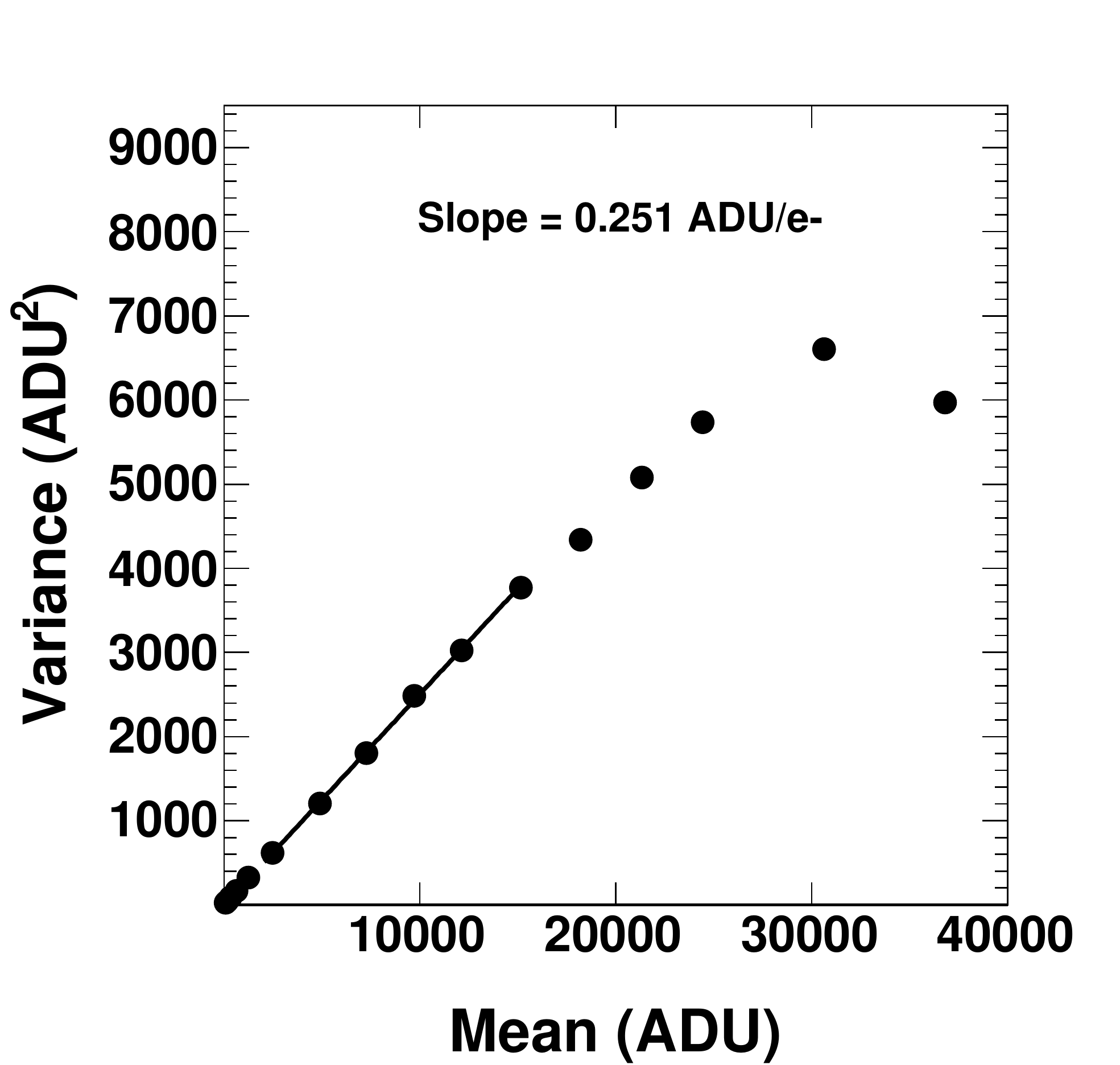}
\caption{Left: Linearity of CCD response for various exposure times.  Data are consistent with $\sim$50ms shutter actuation time.  Exposure times of several tens of seconds ensures linearity. Right:  Variance in counts is likewise linear over the region of interest for exposure times.}
\label{f6}
\end{figure*}

Bench tests were undertaken at Argonne National Laboratory from March to July, 2010, to determine the effectiveness of each individual component and of the entire integrated system.  As already described, vacuum was created by the turbopump and maintained by the getter for nearly a month during the testing period.  Once the LS332 was programmed with the proper set-points and Pt thermocouple resistance values (including the resistance of the readout cables), the temperature was stable to within 0.25K.  With the PANVIEW software running within SISPI, the CCDs were read out and combined into a single FITS image in $\sim$20 seconds.  Other bench tests (e.g., photon transfer curves, see Figure~\ref{f6}) determined the readnoise, dark current, and full well of the CCDs .  At the conclusion of these tests, the PreCam instrument was shipped to CTIO for installation and operation on the C-S telescope.    In the next section, we focus on the PreCam survey operations and data-taking.

\section{Operations and Data-Taking}

\subsection{PreCam Installation and Commissioning on the Curtis-Schmidt}

\begin{figure*}
%% \epsscale{2.5}
%% \plotone{f1.eps}
%% \plottwo{camera_install1.eps}{cameramonsoon.eps}
\plottwo{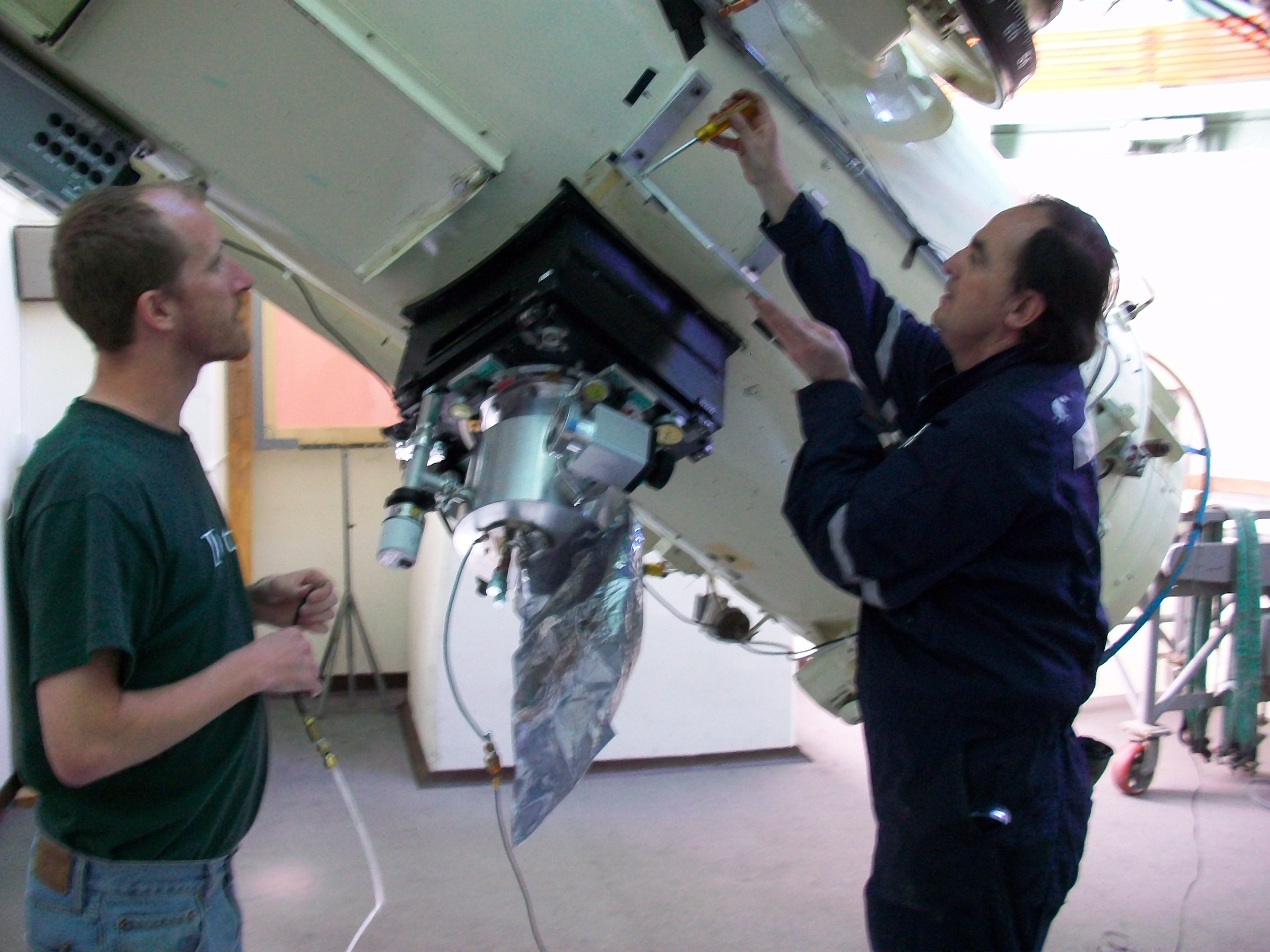}{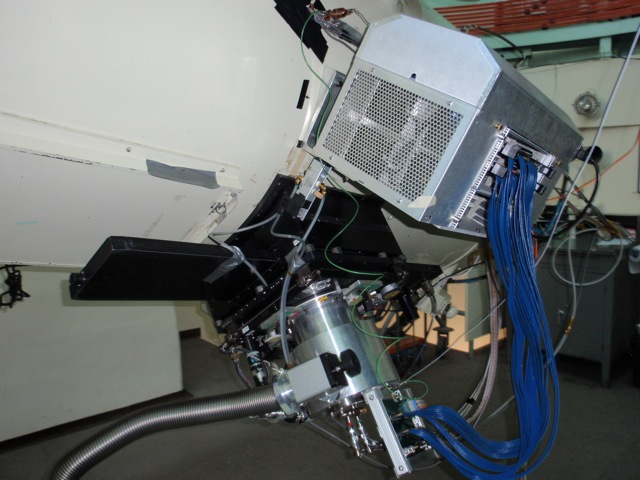}
\caption{Left: Installation of the PreCam instrument on the University of Michigan Curtis-Schmidt telescope at CTIO.  Right: A close-up of the PreCam instrument fully installed, with associated DAQ electronics connected to the Vacuum Interface Board for data readout.}
\label{f7}
\end{figure*}

We were scheduled for 112 nights of telescope time (including installation and testing) through the University of Michigan Department of Astronomy on their Curtis-Schmidt telescope at CTIO.  The C-S telescope is a 0.61 meter aperture telescope of classical Schmidt design, with the CCDs mounted at a Newtonian focus.  Its fast optics and wide field of view make it ideally suited for long-duration wide-field observation campaigns such as the PreCam survey.  Figure~\ref{f7} shows the PreCam instrument as it is being installed on the C-S telescope, along with a close-up view of the instrument.

\begin{figure*}
\epsscale{1.0}
\plotone{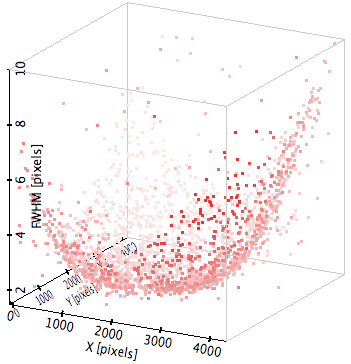}
%%\plotone{fwhmposition.eps}
\caption{Red/dark points show the Full Width at Half Maximum (FWHM) for stars as a function of position on the focal plane.  A flat plane means that a given focus setting will provide comparable FWHM for the entire focal plane, whereas a curved surface means that objects appear larger in some regions of the focal plane.  These data show that, without a field flattener, only a portion of the focal plane is at the best focus for a given image (4096 pixels $\approx$ 1.6$^{\circ}$).}
\label{f8}
\end{figure*}

After installation, testing, and commissioning, we had 64 potential nights of on-sky observing.  Of these, 51 provided useful science data, while the other nights were dedicated to PreCam or C-S engineering tasks (e.g., fixing a broken dome motor), or encountered various hardware or software malfunctions (see Data Processing, Section 4), or, rarely, poor weather conditions.  The testing/commissioning tasks completed during August and September of 2010 included mounting and testing the new camera; installing and testing a new LED-based flat-field system, which was a prototype of the DECal system that is part of the DECam project \citep{DECal}; replacing the C-S's undersized secondary mirror with a new mirror and mount; the subsequent re-collimating of the optics; and interfacing the PreCam instrument with the C-S Telescope Control System (TCS).  After the secondary mirror was replaced it was discovered that the new mirror did not meet requirements and we reinstalled the original secondary and mount.  The final step of the commissioning process involved characterizing the shape of the focal plane and determining the best focus settings of the C-S mirrors.  Although the Schmidt telescope has a curved focal plane, we elected to observe without a field flattener.  Customized software for focus optimization was developed \citep{Sahar}; output showing the curvature of the focal surface is shown in Figure~\ref{f8}.  From these observations, filter- and temperature-dependent focus settings were determined and a procedure for nightly focus determination was developed.

\subsection{The Flat-Field System and PreCam Filters}

\begin{figure*}
\epsscale{2.0}
\plotone{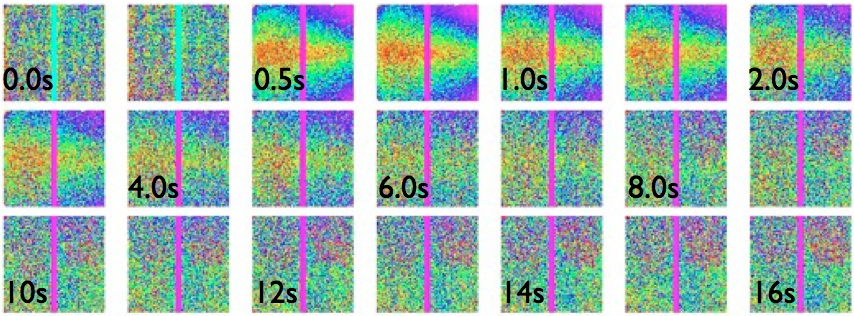}
\caption{Shutter actuation effects as a function of exposure time, from 0 s (top left) to 16 s (lower right).  The nonuniformities clearly disappear by the 8 s image (second row, rightmost image).  This shows that any exposure greater than 8 s---including all PreCam observations---will not be significantly impacted by the shutter actuation time.}
\label{f4}
\end{figure*}

The PreCam flat-field system consists of a screen placed within the C-S dome and illuminated by six sets of LEDs mounted near the top of the C-S telescope.  These LEDs emit light at wavelengths of 470, 740, 905, 970, and 1000 nm, as well as a broad-spectrum ``white'' light.  A seventh LED was included in the flat-field system but was not used for PreCam observations.  These were used to ensure uniformity of the system response to sources of uniform illumination; where variations from uniformity were identified within the observed data, a standard procedure was applied to ``flatten'' the system response so that every pixel of the CCDs effectively would register the same number of counts for a given source intensity.  This also allowed us to measure (and correct for) any possible inaccuracies in collimation of the optics that would cause non-uniform illumination across the focal plane.  The flat-field system is also used to test the effects of the shutter opening/closing time on image uniformity.  Figure~\ref{f4} shows that such non-uniformities are negligible for any image longer than 8 s even without any shutter map corrections; thus, PreCam images are unaffected by the non-zero shutter actuation time, as they are all 10 s or greater in duration.

\begin{figure*}
\epsscale{1.25}
%%\plotone{plotDESandSDSSResponse2.eps}
\plotone{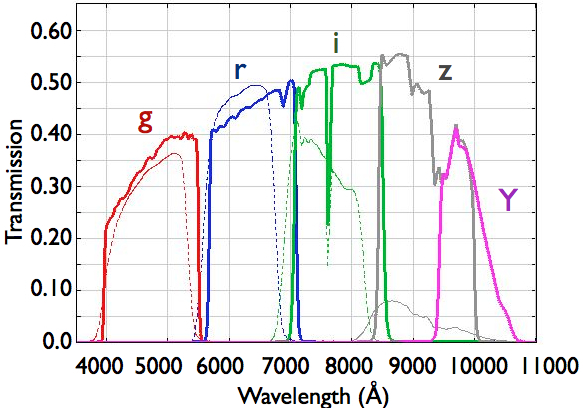}
\caption{Measured transmission as a function of wavelength for PreCam grizY (solid lines) and SDSS griz (dashed lines) filters, convolved with atmospheric transmission and CCD quantum efficiency.  Note particularly the increased effectiveness of the PreCam system at the redder end of the spectrum.}
\label{f9}
\end{figure*}

\begin{figure*}
\epsscale{2.0}
\plotone{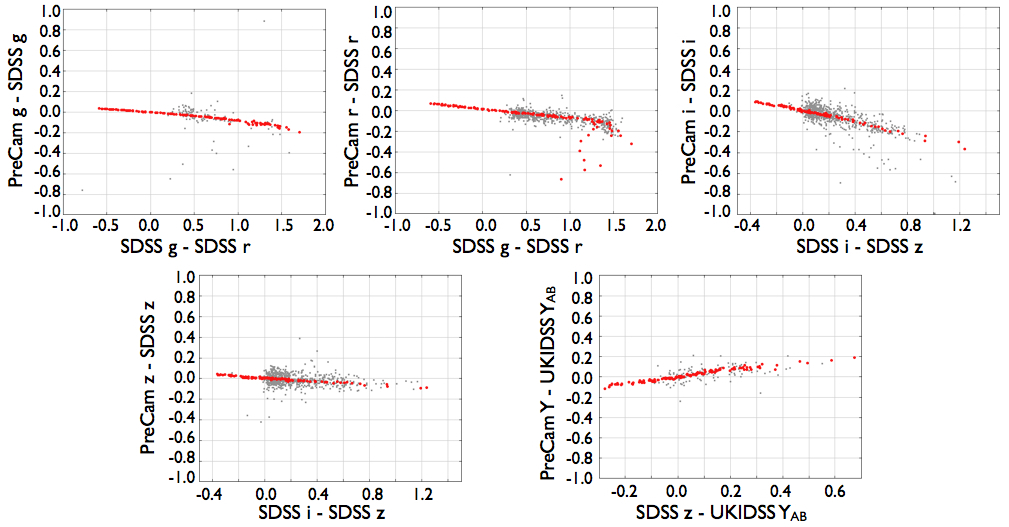}
\caption{PreCam photometry compared to SDSS photometry for g, r, i, z, and Y bands as a function of SDSS/UKIDSS color (grey/light points).  Because of the different filter and CCD response of PreCam compared to SDSS/UKIDSS, a transformation is required to place the observations on identical systems.  With an additional constant (zero-point) offset applied, the linear trends smoothly overlap the Pickles theoretical stellar library (red/dark points).  While Red Giant Branch stars appear among the theoretical points for the r-band, there do not appear to be any such stars among the observed data.  Both the zero-point offset and the linear color terms for each filter are applied in the final comparison of PreCam and SDSS/UKIDSS photometry.}
\label{f10}
\end{figure*}

Scaled-down versions of filters very similar to those used by the DECam were also incorporated into the PreCam instrument, and they performed as expected (hereafter the PreCam filters are referred to with the subscript $pc$).  The measured wavelength-dependent transmission of the filters convolved with atmospheric absorption and CCD quantum efficiency is shown in Figure~\ref{f9}, while Figure~\ref{f10} shows the color terms for a selection of observed stars, relative to the SDSS and UKIDSS filters, compared to those from the Pickles stellar library \citep{Pickles}.  The excellent agreement between the expected and observed performance of the filters is thus confirmed.  A comparison of the PreCam filters to those actually manufactured for and installed on the DECam also will be performed to ensure that any differences between the two sets of filters are accounted for prior to the application of PreCam data to DES calibrations.

\subsection{Observing Strategy and the Observing Tactician}

During primary data-taking (November 2010 to January 2011), each night of observing was preceded by collection of bias (0 s duration) and dark images.  Around sunset (when there is minimal stray light illuminating the interior of the C-S dome), flat-field images were also acquired using the previously-described system.  Once the sky was dark enough (i.e. after astronomical twilight), we made pointing and focus determination observations before proceeding to regular observations.

As part of the PreCam observing procedure, a prototype of the DES Observing Tactician, or ObsTac \citep{ObsTac}, was implemented and tested; this automated target selection and scheduling program was used for the vast majority of the science data-taking, leading to significant improvements in performance that will be carried over to the DES itself.  ObsTac has two primary elements: 1) a database that contains tables of field positions, desired exposure times for each filter, exposures already completed or planned, and other data that influence target selection; and 2) a SISPI service, the ObsTac server, that returns the specification for an observation (pointing, exposure time, filter, and earliest and latest acceptable times of observation) upon receiving a request for an observation at a given time.

When called, the ObsTac server determines if there are any exposures needed with time restrictions (calibration fields, for example), and if there are, it returns the relevant data for such an exposure. Otherwise, it returns the highest priority exposure of the survey that is observable, where ``observable'' takes into account the airmass and phase and separation from the moon. The algorithm used to calculate the priority was modified several times over the course of PreCam observing, but setting time and sky brightness were generally the determining factors. Fields that set first were targeted first, as they may not have been available later in the survey. Exposures in bluer filters were preferred in dark sky conditions, because redder filters can better tolerate brighter sky conditions.  In standard observing with the DES, there will also be additional supporting services---for example, a simple SISPI service will monitor the length of the observing queue, and fill it when necessary through calls to the ObsTac server.

For the PreCam survey, however, not all aspects of the ObsTac infrastructure were used; instead, we performed simulations of the night to be observed, recording which exposures were selected by ObsTac at any given time. This record was then provided to the PreCam observing software, which read and executed the automating observing plan.  Because the plan was generated for the whole night ahead of time, ObsTac could not automatically accommodate changes in observing in real time. In particular, the actual times of observation occasionally differed from those planned by ObsTac, resulting in more large airmass observations than originally expected.  Because airmass-dependent corrections as well as FWHM selection criteria were applied to the observed data, such high-airmass observations are not expected to negatively impact our final results (apart from perhaps reducing the total number of usable standards stars that are incorporated into our catalog).

Based on the ObsTac observing plan, the PreCam observing software interacted with the C-S TCS to move the telescope to the desired observing position, changed the filter to the desired setting, opened and closed the shutter, and interfaced with the DAQ system that recorded and read out each image.  Once each image was completed, the observing script selected the next object in the observing plan and repeated the observing process until all objects were observed, morning twilight prevented further observing, or the observer manually interrupted the observing sequence.  Guided by this automated process, we obtained 11020 images during the PreCam survey.  The majority of these images were located along SDSS Stripe 82 and throughout the eastern half of the DES footprint, though we also obtained some images outside that region.  PreCam survey images contain just under 10 million identified objects, though this includes multiple observations of the same objects.  All observations are divided into three distinct categories: calibration exposures (biases, dark frames, dome flats), standard fields (10 s images of known bright standards), and target fields (including both known and candidate standard stars).  Standard observations were repeated in every filter once every hour throughout the observing nights in order to ensure the internal consistency of the PreCam dataset.  Furthermore, these observations were used to determine the average zero-point offset for data taken with each filter on each night.  With these zero-points offsets applied, the photometric accuracy is greatly improved (see Results, below).

\subsection{The Quick-Reduce Pipeline}

During observations, a prototype of the Quick Reduce (QR) pipeline developed for the DECam by the DES-Brazil team was deployed at CTIO and adapted to handle data from the PreCam survey.  The QR pipeline carries out the following steps: 1) automatically reads the most recent image stored on disk; 2) applies standard corrections to this image (overscan, bias subtraction and flat-field correction); 3) extracts a catalog using Source Extractor \citep{SEx}; 4) computes the mean of the measured FWHM and evaluates the distortion of the Point Spread Function (PSF) of the sources classified as stars on each CCD; 5) compares these values to user-specified values to determine whether the image complies with the quality criteria established; and 6) produces a set of web pages summarizing the results and showing a JPEG representation of the images reduced. The QR pipeline also allows displaying the raw and reduced fits files using, for instance, DS9.  While the use of the QR pipeline during PreCam observing was limited, the experience was extremely valuable in the development and testing of the final system that has been incorporated into SISPI for use with the DES.

At the end of each night, the raw PreCam data were also transferred to the DES tertiary data storage site in Rio de Janeiro, Brazil, and subsequently to Fermilab using a tool named Bits Around the World (BAW) developed by
the DES-Brazil group as a substitute for the official CTIO Data Transfer System (DTS), which was still under development at the time.  These data were subsequently used in the development and testing of a pipeline for regular, automated, complete reduction of PreCam and (soon-to-be-acquired) DES data.  Although the pipeline was not used to reduce all of the accumulated PreCam data, it was applied to several nights of data, demonstrating that approximately eight hours of data can be successfully reduced and the results transferred during the following daytime period, enabling a rapid feedback mechanism for guiding observations on subsequent nights, a critical component of the DES Observing Strategy.

\begin{figure*}
\epsscale{2.0}
\plotone{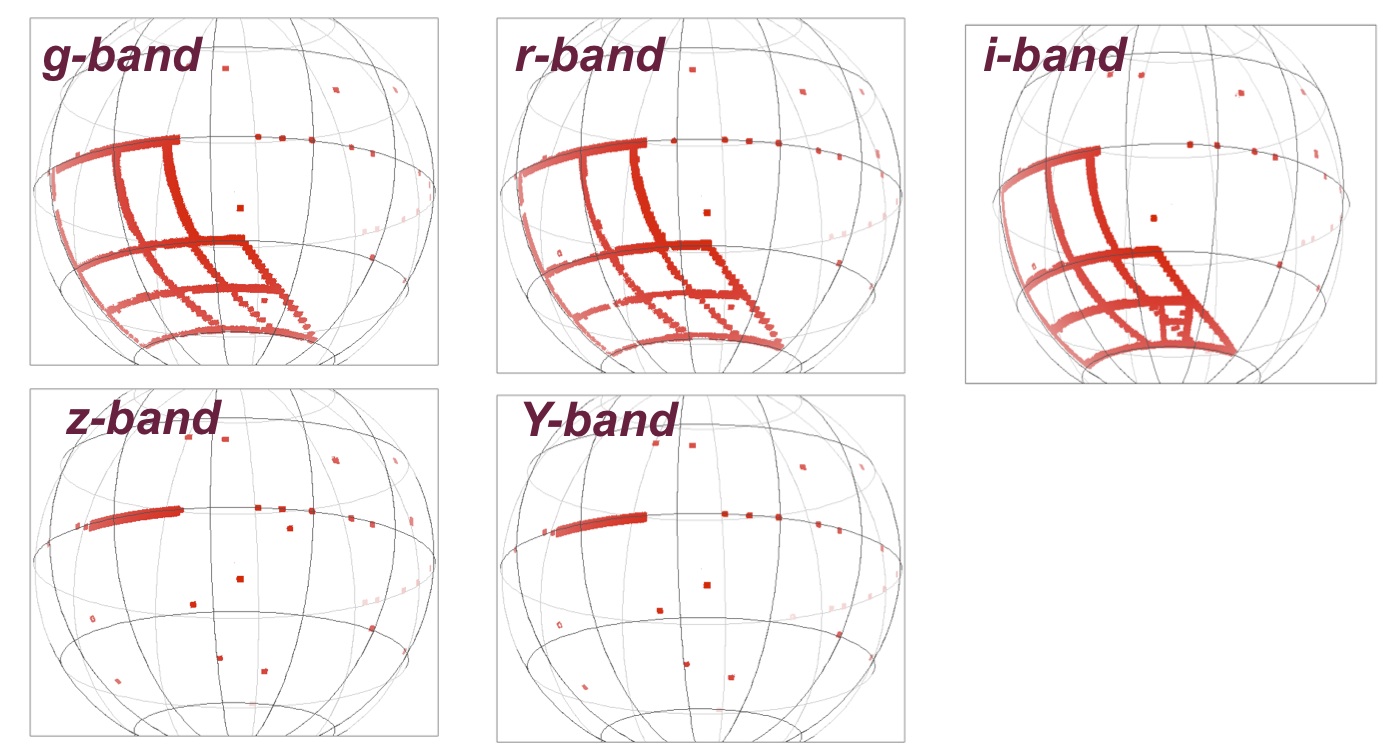}
\caption{Coverage maps showing the areas of the PreCam grid observed for each of the five filters during Season One of operations.  Points outside of the grid are known standard star fields also observed during the PreCam survey.  We are currently considering a second season of observations to complete the grid in all five filters.}
\label{f11}
\end{figure*}

Table~\ref{tab2} shows the total number of science and calibration images obtained per filter (after initial data quality selection), along with the number of stars identified in these images, while Figure~\ref{f11} shows the completed coverage map overlaid on the DES footprint, as well as filter-specific coverage maps.

\begin{table*}
\begin{center}
\begin{tabular}{lcccc}
\hline
\tablehead{1}{Filter} & \tablehead{1}{Target Field Exposure Duration (s)} & \tablehead{1}{Number$_\mathrm{Target Exposures}$} & \tablehead{1}{Number$_\mathrm{Std. Exposures}$} & \tablehead{1}{Number$_\mathrm{Stars}$} \\
\hline
g$_\mathrm{pc}$ & 36 & 1888 & 700 & 959,085 \\
r$_\mathrm{pc}$ & 51 & 1868 & 700 &  1,881,279 \\
i$_\mathrm{pc}$ & 65 & 3152 & 700 & 3,881,716 \\
z$_\mathrm{pc}$ & 162 & 269 & 700 & 1,816,290 \\
Y$_\mathrm{pc}$ & 73 & 343 & 700 & 890,181 \\
\hline
\end{tabular}
\caption{Number of exposures of target fields and standard fields (observed with 10 s exposures), as well as the number of stars identified from both types of exposures in the PreCam observations, for each PreCam (pc) filter.}
\label{tab2}
\end{center}
\end{table*}

\section{Preliminary Data Processing}

Once the data were collected, we performed standard bias correction and flat-fielding of all the images (at -100$^{\circ}$C, the dark current was negligible and thus was not subtracted from the images), using master bias and master flat-field images that were derived from all useful images of these types.  After these standard steps, further processing was required due to unique circumstances confronted during PreCam observations.  The full details of the final processing and analysis steps will be detailed in a subsequent publication \citep{Allam}; here we describe the preliminary analysis applied to a representative sample ($\sim$10$\%$, described in Table~\ref{tab3}) of the data collected.

\begin{table}
%% \begin{table*}
%% \begin{center}
\begin{tabular}{lcr}
\hline
\tablehead{1}{Date (UT)} & 
\tablehead{1}{Filter(s) Used} &
\tablehead{1}{N$_\mathrm{Target Exposures}$}
\\
\hline
21011215 & i, z & 40, 29\\
20110107 & g, z & 194, 3\\
20110108 & r, z & 56, 10\\
20110112 & i, z & 19 , 19\\
20110113 & i & 174 \\
20110117 & i & 207 \\
\hline
\end{tabular}
\caption{Observation log for the 10\% subsample analyzed here, detailing the number of target exposures taken, the date, and the filter used.}
\label{tab3}
%% \end{center}
%% \end{table*}
\end{table}

\subsection{PreCam-Specific Image Processing: Streaking and Banding Corrections}

\begin{figure*}
\epsscale{2.0}
\plotone{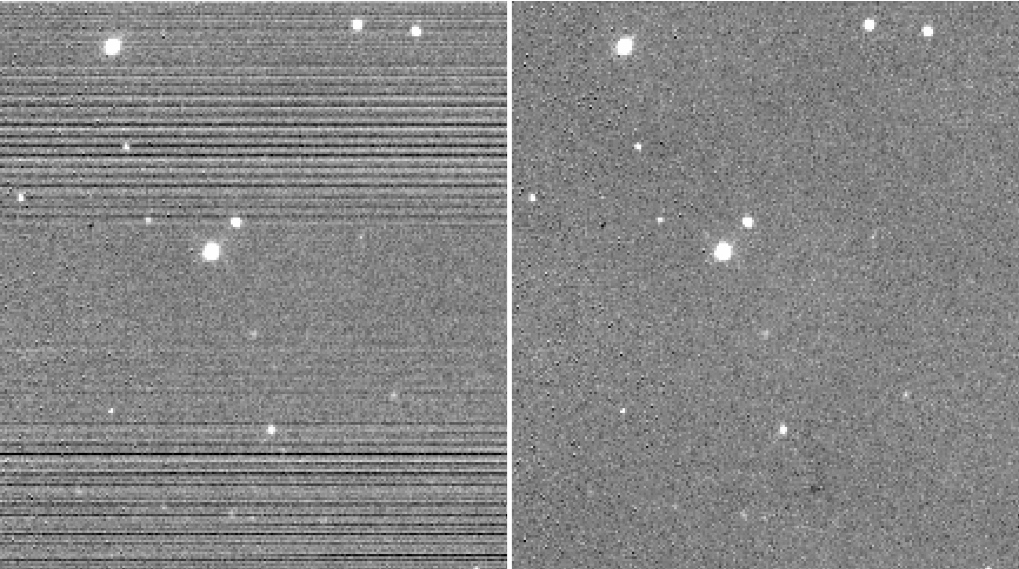}
\caption{Left: A representative PreCam image with significant streaking that would interfere with the photometric determination of the stars in that field.  Right: The same image after streaking removal, exhibiting desired behavior for both background regions and stars.}
\label{f12}
\end{figure*}

One significant problem encountered during data-taking was horizontal streaking and banding within the images.  This was eventually traced to microscopic damage within the cables connecting the VIB to the Monsoon crate.  Design differences preclude this problem from occurring on the DECam; meanwhile, in preparation for future PreCam data-taking, we have prepared a repaired cable as well as a strain-relief system that prevents the weight of the cable from damaging the sensitive connections as the telescope is moved.  PreCam data that have already been collected were scanned for the presence of streaking or banding; nearly 40$\%$ of the images showed some signs of streaking.  Because this induced common-mode noise was observed to be cyclic and to exhibit the same pattern in each amplifier for each CCD, a straightforward software algorithm comparing the pixel-by-pixel counts in the four amplifiers permitted the identification (and removal) of the banding and streaking effects (see Figure~\ref{f12} for an example of an image with significant streaking, and the subsequent corrected image).  Photometric and astrometric accuracy of the vast majority of these images was restored with this process; only 3.6$\%$ of the images were finally considered ``unrecoverable'' due to the presence of significant and uncorrectable streaking.  Meanwhile, the residual uncertainty in photometric accuracy introduced by this process is small (mean variation of 0.01 mag) for any given image; therefore, thousands of additional images were retained in our dataset, and ultimately will contribute to the final PreCam Standard Star Catalog.

\subsection{PreCam-Specific Image Processing: Shutter Corrections}

Repeated actuation of the shutter led to failure of the shutter blades on several occasions.  Ultimately, some of the blades completely broke, thereby preventing the shutter from fully closing (and obviously impeding our data-taking efforts).  Prior to complete failure, however, the shutter blades would occasionally ``stick'' in a partially-open configuration.  As with the streaking and banding problems, DECam's design differences prevent such a problem from occurring during the Dark Energy Survey.  Within the PreCam dataset, only a small fraction ($\sim$3.5$\%$) of the images were identified as suffering from shutter problems, and the vast majority of these are still useful for determining stellar photometry, as a) only a small region of the full image is affected, and b) local background-subtraction algorithms can correct for the increased noise in these regions with negligible degradation of the resulting photometric accuracy.  Additionally, the shutter actuation system has been modified with a slow-release valve for the compressed gas that extends the lifetime of the shutter blades by approximately a factor of three relative to the original shutter system.  While this may not completely eliminate shutter problems during future operations, it improves shutter performance considerably and, coupled with the near-real-time image analysis provided by the QR system to identify broken or stuck shutter blades, will prevent significant loss of observing time during any future PreCam observing campaigns.  The effect of the increased shutter closing time upon the uniformity of focal plane illumination is still being quantified, but because it only increases the blade actuation time by a factor of $\sim$2, it is not expected to impact PreCam exposures of 10 s or longer.

\subsection{PreCam-Specific Image Processing: Illumination Correction}

\begin{figure*}
\epsscale{2.0}
\plottwo{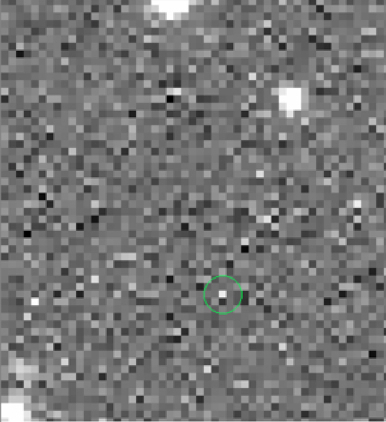}{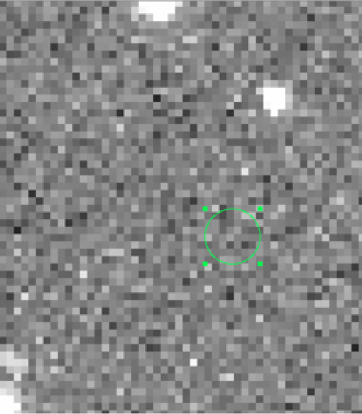}
\caption{Left: A segment of a representative PreCam image before illumination correction, showing vertically-oriented light and dark spots (``dipoles'') inside the green circle.  Right: The same region of the image after Illumination Correction, showing that the dipoles have been removed.}
\label{f13}
\end{figure*}

Finally, minor electronics issues associated with the CCDs and readout electronics arose during data processing.  Specifically, ``dipoles'' became apparent in the partially-processed images.  These occur when charge is improperly separated between (vertically) adjacent pixels, and one pixel is observed with a significantly greater background value than its immediate neighbor.  These and other spurious variations are eliminated both in the production versions of the DES hardware and in the upgraded PreCam hardware; for all PreCam data already obtained these variations are eliminated by Illumination Correction processing (see Figure~\ref{f13}).  Illumination Correction is effectively an additional application of flat-field corrections, incorporating a pixel-by-pixel multiplicative factor based on the mean of all target field exposures for each filter during a given night of observing (as opposed to the initial flat-fielding, which was based on master flat images constructed from dedicated dome flat exposures taken throughout the duration of the survey).  Instrument-induced variation between pairs of adjacent pixels is corrected in this fashion, from $\sim$16$\%$ to a level consistent with normal background variation, and any stellar profiles affected by dipoles are marginally improved (due to more accurate distribution of stellar photon counts).  However, the improvement to the photometry of known standard stars is negligible solely by removing the dipoles, because (in general) the values for both pixels was already incorporated into any stellar photometry measurements, since all the light that came from a star (even that portion of it translated by one pixel) was still associated with that star.  However, for the longer exposures (i$_\mathrm{pc}$-, z$_\mathrm{pc}$-, and Y$_\mathrm{pc}$-band images), Illumination Correction removed significant additional variations in illumination not corrected by the initial flat-fielding process, thus improving the final photometric accuracy of the stars found in images in those three bands.  Because the photometric accuracy of stars in the g$_\mathrm{pc}$- and r$_\mathrm{pc}$-band images was not improved by the Illumination Correction, this step was not applied to images taken in these filters.

\subsection{Star Flat Corrections and Catalog Generation}

\begin{figure*}
\epsscale{1.0}
\plotone{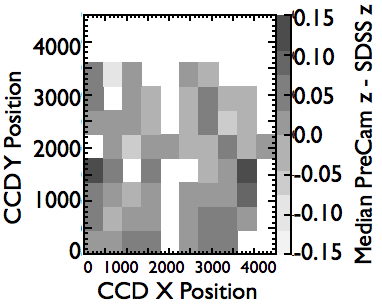}
\caption{A representative star flat image derived from known standard star observations for a given filter on a given night.  When applied in a manner similar to the standard flat-field images, the star flat was shown to further correct the residual variations in photometric accuracy.}
\label{f14}
\end{figure*}

After the streaking/banding, shutter, and illumination variation issues are resolved, as a final processing step, residual instrumental effects are removed by means of applying a star flat correction (see Figure~\ref{f14}).  These star flat images are created by comparing, over the course of a night, the magnitudes of known standard stars with their previously measured magnitudes from, e.g., SDSS or other catalogs.  This yields a fine-scale (of order 100 sub-regions per image) zero-point correction that is applied to each image in addition to the global zero-point for the entire image. After this processing, a catalog is developed for each image using Source Extractor.  The preliminary PreCam Southern Star Catalog is derived from this catalog, after the application of additional selection criteria; specifically, the Source Extractor-determined Stellarity $>$0.95 and FWHM $<$4 pixels (based upon a 12 arcsec diameter circular aperture).  In the next section we describe results from the $\sim$10\% subsample of the PreCam stellar catalog to which we have applied this preliminary analysis.

\section{Results}

\begin{figure*}
\epsscale{2.0}
\plotone{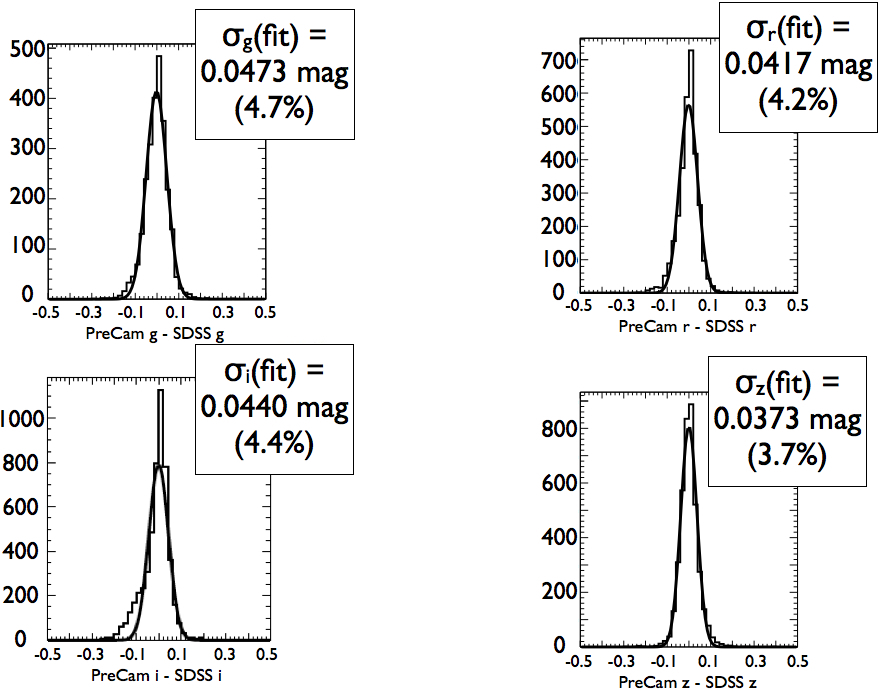}
\caption{Distribution of photometric accuracy of a representative set of stars observed in g$_\mathrm{pc}$-, r$_\mathrm{pc}$-, i$_\mathrm{pc}$-, and z$_\mathrm{pc}$-band images relative to SDSS measurements.  Nearly all stars measured here are between 14th and 17th magnitude.  The results show that the fit to the preliminary single-epoch accuracy has a sigma between 3$\% $and 5$\%$ for these four filters.  The final processing and analysis algorithm based on the averaging of multiple images of each field is expected to improve these results to between 1$\%$ and 2$\%$.}
\label{f15}
\end{figure*}

\begin{figure*}
\epsscale{1.75}
\plotone{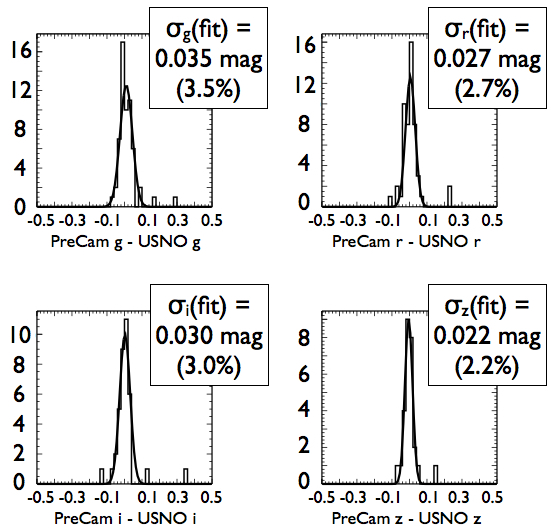}
\caption{Distribution of photometric accuracy of a representative set of stars observed in g$_\mathrm{pc}$-, r$_\mathrm{pc}$-, i$_\mathrm{pc}$-, and z$_\mathrm{pc}$-band images relative to the Equatorial and Southern extensions of the u$^{\prime}$g$^{\prime}$r$^{\prime}$i$^{\prime}$z$^{\prime}$ standard star catalog observed with the U.S. Naval Observatory (USNO) 40 inch telescope.  Nearly all stars measured here are between 10th and 15th magnitude.  The results show that the fit to the preliminary single-epoch accuracy has a sigma between 2$\%$ and 3.5$\%$ for these four filters.  The final processing and analysis algorithm based on the averaging of multiple images of each field is expected to improve these results to between 1$\%$ and 2$\%$.}
\label{f16}
\end{figure*}

Preliminary measurements of target candidate standard stars in the PreCam images are compared to the 2MASS Point Source catalog \citep{2MASS} for astrometry, while for photometry they are compared to the SDSS DR7 catalog \citep{Sloan2} as well as the Equatorial and Southern extensions of the u$^{\prime}$g$^{\prime}$r$^{\prime}$i$^{\prime}$z$^{\prime}$ catalogs \citep{South2}.  For the astrometry, we use a set of python scripts to perform a ``first pass'' fit, which is good to $\sim$2 arcseconds (RMS).  We are currently refining these scripts and, with proper treatment of the field curvature and using independent astrometric solutions for the two PreCam CCDs, we have achieved astrometric accuracy good to $\sim$0.3 arcseconds (RMS), or about $\frac{1}{5}$ of a pixel.  This is sufficient to provide unique identification of any stars that will be included in the final version of the PreCam Standard Star Catalog.  For the photometry, existing standard stars are matched if the PreCam-determined position was within 3 arcseconds of the known position, and then nightly zero-point offsets and airmass-dependent color corrections are applied to the matched stars prior to determining the final photometric accuracy.  A representative sample of the results taken from $\sim$10$\%$ of the observations, or six nights out of the total of 51, shows that PreCam measures stellar photometry with a mean accuracy of 3-5$\%$ (depending on filter) relative to SDSS results, and better than that ($\sim$2$\%$) for Southern u$^{\prime}$g$^{\prime}$r$^{\prime}$i$^{\prime}$z$^{\prime}$ stars brighter than magnitude 15 (see Figures~\ref{f15} and ~\ref{f16}).  For the final PreCam catalog with improved data analysis and averaging of multiple observations of each field, our photometry is expected to achieve the required 2$\%$ accuracy \citep{Allam}.

\begin{figure*}
\epsscale{1.5}
\plotone{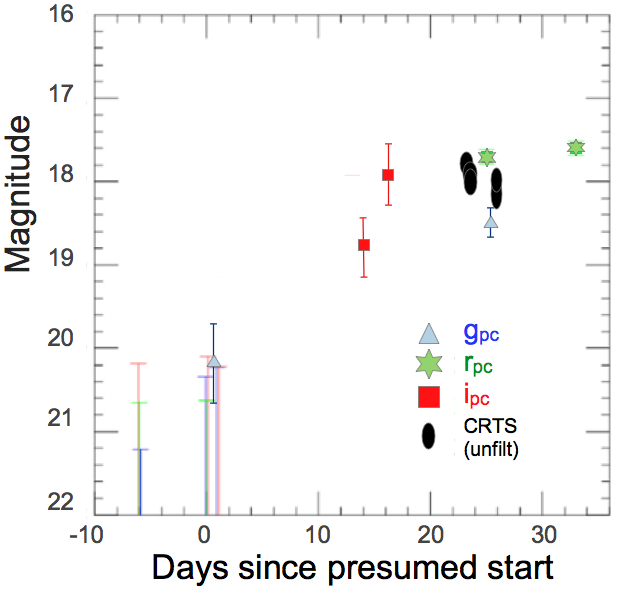}
\caption{PreCam (pc) and Catalina Real-Time Transient Survey (CRTS) lightcurves of supernova SN2010lr, showing pre-detection limits and post-detection magnitudes.  Uncertainties in the PreCam measurements are estimated by taking the difference between the Source Extractor-determined MAG\_AUTO and MAG\_APER (with an aperture of 10 arcsec).  The presumed start around December 5 is inferred from the later observations of the CRTS, and the lines extending to the lower edge of the plot on or before this date are pre-detection (5$\sigma$) upper limits set by PreCam.  As described in the text, the first detection used a distinct analysis procedure with only a 1.5$\sigma$ detection threshold, thus we expect it to be significantly less accurate than any of the photometry results comprising our standard star catalog.}
\label{f17}
\end{figure*}

In addition to the production of a preliminary standard star catalog based upon previous standards, PreCam provides a preview of the science that will be conducted with the DES.  SN2010lr, a Type Ia supernova discovered by the Catalina Real-Time Transient Survey, or CRTS \citep{Drake}, was observed by PreCam as well, and we show a preliminary light curve of the SN in Figure~\ref{f17}.  The observed magnitude near 18 is in agreement with photometric measurements from the CRTS.  Spectroscopic measurements by \citep{Prieto} lead these authors to claim that this SN is most similar to SN1998bu with peak magnitude occurring around December 20 (Day 15 of the figure).  Finally, it is worth noting that SN2010lr was seen (after the fact) in PreCam data from 2010-12-05, a date well before that of the initial CRTS discovery image.  This initial detection was made at a magnitude significantly below that for which the PreCam survey is designed, thus the standard analysis procedures described above were not applied; instead, after data processing a single cut using Source Extractor's \tt{DETECT\_THRESH}\rm~of 1.5 was made to generate the object catalogs that include the supernova.  Therefore, we expect the photometric accuracy of this SN measurement to be significantly worse than for any of the standard stars in our catalog; nevertheless, the clear detection of the SN in the PreCam data is an encouraging sign of what will be accomplished with the significantly larger and more sensitive Dark Energy Survey.

\section{Discussion}
PreCam, a precursor observational campaign for calibration of the Dark Energy Survey, was designed to provide accurate photometric and astrometric measurements of stars in a sparse grid throughout the DES footprint, while the PreCam instrument also served as a testbed for various components of the Dark Energy Camera.  We successfully completed observations of roughly half of the desired target fields comprising the PreCam survey grid by early 2011, and preliminary results based on the analysis of a subset of the data show that we are approaching the stringent requirements for calibration of the DES.  During construction and operations we tested the pre-production DES DAQ system, (spare) CCDs for the DECam, ObsTac and the QR system, as well as various other components.  As described above, many different issues were overcome during the commissioning and observations of PreCam, including a secondary mirror of insufficient optical quality, banding and streaking within the data, shutter failures, and data transport issues.  Nevertheless, we were able to collect data during 51 out of 64 possible nights ($\sim$80\% duty cycle), and the vast majority of those data (well in excess of 90\%) provides us with useful measurements of thousands of previously-measured standard stars and up to a hundred thousand potential new standard stars to a magnitude limit if i$_{pc}$$\sim$17.  All of these will be incorporated into the forthcoming PreCam Standard Star Catalog and used as inputs to the DECam calibration procedures.  Additional observations with the PreCam instrument to complete the proposed grid of new standard stars, using the Curtis-Schmidt or another telescope, are under consideration but are not approved at this time.

\section{Conclusions}

In an effort to provide a catalog of standard stars for improved calibration of the Dark Energy Survey, we designed, built, tested, installed, commissioned, and operated the PreCam instrument to collect observations over 51 days in 2010 and 2011.  We performed PreCam survey observations in a sparse grid throughout the Dark Energy Survey footprint; photometric accuracy as compared to the SDSS catalog is between 3.0\% and 5.0\% for the sample of data described in this paper, and 2\% as compared to the brighter stars of the u$^{\prime}$g$^{\prime}$r$^{\prime}$i$^{\prime}$z$^{\prime}$ standard star catalog extensions.  Final analysis of these data should allow us to reach the 2$\%$ photometric accuracy for nearly a hundred thousand stars suitable for DES standard star calibrations, and perhaps even obtain the desired 1$\%$ accuracy.  The full PreCam Southern Hemisphere Standard Star Catalog will be released after complete analysis of the full PreCam dataset \citep{Allam}, but application of the standard star calibrations to the DES data will begin with the first light of DES, currently scheduled for September 2012.

%%%%%%%%%%%%%%%%%%%%%%%%%%%%%%%%%%%%%%%%%%%%%%%%
%% BACKMATTER
%%%%%%%%%%%%%%%%%%%%%%%%%%%%%%%%%%%%%%%%%%%%%%%%

\acknowledgments

The submitted manuscript has been created by UChicago Argonne, LLC,
Operator of Argonne National Laboratory (``Argonne''). Argonne, a
U.S. Department of Energy Office of Science laboratory, is operated under
Contract No. DE-AC02-06CH11357. The U.S. Government retains for itself,
and others acting on its behalf, a paid-up nonexclusive, irrevocable
worldwide license in said article to reproduce, prepare derivative
works, distribute copies to the public, and perform publicly and
display publicly, by or on behalf of the Government.

This paper has gone through internal review by the DES collaboration.
Funding for the DES Projects has been provided by the U.S. Department of Energy, the U.S. National Science Foundation, the Ministry of Science and Education of Spain, the Science and Technology
Facilities Council of the United Kingdom, the Higher Education Funding Council for England, the National Center for Supercomputing Applications at the University of Illinois at
Urbana-Champaign, the Kavli Institute of Cosmological Physics at the University of Chicago,
Financiadora de Estudos e Projetos, Funda\c{c}\~{a}o Carlos Chagas Filho de Amparo \`{a} Pesquisa do Estado do Rio de Janeiro, Conselho Nacional de Desenvolvimento Cient\'{i}fico e
Tecnol\'{o}gico (CNPq - Brazil) and the Minist\'{e}rio da Ci\^{e}ncia, Tecnologia, e Inova\c c\~ao (MCTI - Brazil), the Deutsche Forschungsgemeinschaft and the Collaborating Institutions in the Dark Energy Survey.

The Collaborating Institutions are Argonne National Laboratory, the University of California at Santa Cruz, the University of Cambridge, Centro de Investigaciones Energeticas,
Medioambientales y Tecnologicas-Madrid, the University of Chicago, University College London, DES-Brazil, Fermilab, the University of Edinburgh, the University of Illinois at Urbana-Champaign,
the Institut de Ciencies de l'Espai (IEEC/CSIC), the Institut de Fisica d'Altes Energies, the Lawrence Berkeley National Laboratory, the Ludwig-Maximilians Universit\"at and the associated
Excellence Cluster Universe, the University of Michigan, the National Optical Astronomy Observatory, the University of Nottingham, the Ohio State University, the University of Pennsylvania, the
University of Portsmouth, SLAC, Stanford University, the University of Sussex, and Texas A\&M University.  Fermilab is operated by Fermi Research Alliance, LLC under Contract No. DE-AC02-07CH11359 with the United States Department of Energy.  SLAC is operated by Stanford University under contract No. DE-AC02-76SF00515.

The authors recognize the efforts of the Fermilab engineers who were instrumental in designing, producing, and testing pre-production DES detector hardware for the PreCam instrument, particularly Theresa Shaw and Walter Stuermer.  We likewise thank former members of the DES-Brazil group, Leandro Martelli and Bruno Rossetto, for their support of the PreCam efforts. The authors are also grateful for the significant contributions of the professional staff of Cerro Tololo Interamerican Observatory, including Marco Bonati, Gale Brehmer, Jorge Briones, Oscar Saa, and the TELOPS staff, especially during installation, commissioning, and operation of the PreCam instrument.

The U-M Curtis-Schmidt telescope is dedicated to optical observations of space debris, in a program funded by grants to the University of Michigan from the NASA Orbital Debris Program Office.  P. Seitzer (Principal Investigator) thanks the Office for their long term and continuing support.  In particular the Debris Program funded upgrades to the Curtis-Schmidt which made an automated survey project like PreCam possible.

This paper makes use of data from the Sloan Digital Sky Survey.  Funding for the SDSS and SDSS-II has been provided by the Alfred P. Sloan Foundation, the Participating Institutions, the National Science Foundation,
the U.S. Department of Energy, the National Aeronautics and Space Administration, the Japanese Monbukagakusho, the Max Planck Society, and the Higher Education Funding Council for England. The SDSS Web Site isÊhttp://www.sdss.org/ [www.sdss.org].

The SDSS is managed by the Astrophysical Research Consortium for the Participating Institutions. The Participating Institutions are the American Museum of Natural History, Astrophysical Institute Potsdam, University of
Basel, University of Cambridge, Case Western Reserve University, University of Chicago, Drexel University, Fermilab, the Institute for Advanced Study, the Japan Participation Group, Johns Hopkins University, the Joint Institute
for Nuclear Astrophysics, the Kavli Institute for Particle Astrophysics and Cosmology, the Korean Scientist Group, the Chinese Academy of Sciences (LAMOST), Los Alamos National Laboratory, the Max-Planck-Institute for
Astronomy (MPIA), the Max-Planck-Institute for Astrophysics (MPA), New Mexico State University, Ohio State University, University of Pittsburgh, University of Portsmouth, Princeton University, the United States Naval Observatory, and the University of Washington.

Finally, this publication makes use of data products from the Two Micron All Sky Survey, which is a joint project of the University of Massachusetts and the Infrared Processing and Analysis Center/California Institute of Technology, funded by the National Aeronautics and Space Administration and the National Science Foundation.

%% {\it Facilities:} \facility{Blanco}.

%% \bibliographystyle{plainnat}

\clearpage
\end{document}